\pdfoutput=1
\RequirePackage{ifpdf}
\ifpdf 
\documentclass[pdftex]{sigma}
\else
\documentclass{sigma}
\fi

\usepackage{mathtools}

\begin{document}


\renewcommand{\PaperNumber}{073}

\FirstPageHeading

\ShortArticleName{Direct Connection between the $\text{R}_{\text{II}}$ Chain and the ND-MKdV Lattice}
\ArticleName{Direct Connection between the $\text{R}_{\text{II}}$ Chain \\
and the Nonautonomous Discrete\\
Modif\/ied KdV Lattice}

\Author{Kazuki MAEDA and Satoshi TSUJIMOTO}
\AuthorNameForHeading{K.~Maeda and S.~Tsujimoto}

\Address{Department of Applied Mathematics and Physics, Graduate School of Informatics,\\
Kyoto University, Kyoto 606-8501, Japan}

\Email{\href{mailto:kmaeda@amp.i.kyoto-u.ac.jp}{kmaeda@amp.i.kyoto-u.ac.jp},
\href{mailto:tujimoto@i.kyoto-u.ac.jp}{tujimoto@i.kyoto-u.ac.jp}}

\ArticleDates{Received September 20, 2013, in f\/inal form November 22, 2013; Published online November 26, 2013}

\Abstract{The spectral transformation technique for symmetric $\text{R}_{\text{II}}$ polynomials is developed.
Use of this technique reveals that the nonautonomous discrete modif\/ied KdV (nd-mKdV) lattice is directly
connected with the $\text{R}_{\text{II}}$ chain.
Hankel determinant solutions to the semi-inf\/inite nd-mKdV lattice are also presented.}


\Classification{37K35; 37K60; 42C05}

\section{Introduction}

In the theory of integrable systems, orthogonal polynomials play an important role.
In particular, the spectral transformation technique yields various integrable systems and particular
solu\-tions~\cite{papageorgiou1995opa,spiridonov1995ddt,spiridonov1997dtv}.
The spectral transformation for orthogonal polynomials is a~mapping from an orthogonal polynomial sequence
to another orthogonal polynomial sequence.
We can view the three-term recurrence relation and the spectral transformation for orthogonal polynomials
as a~Lax pair, where the compatibility condition induces an integrable system.
Furthermore, the determinant structure of orthogonal polynomials allows us to derive particular solutions
to the associated integrable system.
During the last f\/ifteen years, many researchers have extended this technique to generalized
(bi)orthogonal functions and have exploited novel integrable systems that have rich
properties~\cite{adler1999pls,adler1997sop,adler2002tvp,kharchev1997frt,miki2012dst,miki2011cbp,mukaihira2002sfo,
mukaihira2004dsr,tsujimoto2002dtl, tsujimoto2010dso,tsujimoto2000msd,vinet1998icb}.
In this paper, we will extend the spectral transformation technique for symmetric orthogonal polynomials
and the associated discrete integrable system to symmetric $\text{R}_{\text{II}}$ polynomials.
Our motivation comes from applications of discrete integrable systems to numerical algorithms.
It is well-known that the discrete integrable system associated with orthogonal polynomials is the
nonautonomous discrete Toda (nd-Toda) lattice and that its time evolution equation is the same as the
recurrence relation of the dqds algorithm~\cite{fernando1994asv}, a~fast and accurate eigenvalue or
singular value algorithm.
Similarly, the discrete integrable system associated with symmetric orthogonal polynomials is the
nonautonomous discrete Lotka--Volterra (nd-LV) lattice, which can compute singular
values~\cite{iwasaki2004aot}.
By using the spectral transformation technique, we can easily derive a~direct connection between the
nd-Toda lattice and the nd-LV lattice.
This connection was used to develop the mdLVs algorithm, which is an improved version of the singular value
algorithm based on the nd-LV lattice~\cite{iwasaki2006acs}.

Recently, the authors have been developing a~generalized eigenvalue algorithm based on the $\text{R}_{\text{II}}$
chain~\cite{maeda2013gea}.
Since the $\text{R}_{\text{II}}$ chain is associated with $\text{R}_{\text{II}}$ polynomials, a~generalization of orthogonal polynomials from
the point of view of Pad\'e approximation or the eigenvalue
problem~\cite{ismail1995goa,mukaihira2006dsn,spiridonov2000stc}, the proposed algorithm has good properties
similar to the dqds algorithm.
These studies motivate us to f\/ind the discrete integrable system associated with symmetric $\text{R}_{\text{II}}$
polynomials and its connection with the $\text{R}_{\text{II}}$ chain.
The derived discrete integrable system may become the basis for developing good numerical algorithms for
generalized eigenvalue problems or Pad\'e approximations, for example.

This paper is organized as follows.
In Section~\ref{sec:derivation-nd-toda}, we brief\/ly recall the derivation of the nd-Toda lattice
and the nd-LV lattice from the theory of spectral transformations for ordinary and symmetric orthogonal
polynomials, respectively.
We also review the direct connection (Miura transformation) between the nd-Toda lattice and the nd-LV
lattice.
In Section~\ref{sec:derivation-rii-chain}, we extend the framework presented in
Section~\ref{sec:derivation-nd-toda} to $\text{R}_{\text{II}}$ polynomials.
We then demonstrate that the spectral transformations for symmetric $\text{R}_{\text{II}}$ polynomials give rise to the
nonautonomous discrete modif\/ied KdV (nd-mKdV) lattice.
Particular solutions to the semi-inf\/inite nd-mKdV lattice and the direct connection between the $\text{R}_{\text{II}}$
chain and the nd-mKdV lattice are also derived.
Section~\ref{sec:concluding-remarks} is devoted to concluding remarks.

\section{Derivation of the nd-Toda lattice and the nd-LV lattice}
\label{sec:derivation-nd-toda}

\subsection{Orthogonal polynomials and the nd-Toda lattice}

Monic orthogonal polynomials are def\/ined by a~three-term recurrence relation in the form
\begin{gather}
\phi^{k,t}_{-1}(x)\coloneqq0,
\qquad
\phi^{k,t}_0(x)\coloneqq1,
\nonumber
\\
\phi^{k,t}_{n+1}(x)\coloneqq\big(x-a^{k,t}_n\big)\phi^{k,t}_n(x)-b^{k,t}_n\phi^{k,t}_{n-1}(x),
\qquad
n=0,1,2,\dots,
\label{eq:trr-op}
\end{gather}
where $a^{k, t}_n \in \mathbb R$, $b^{k, t}_n \in \mathbb R\setminus \{0\}$, and $k, t \in \mathbb Z$ indicate
discrete time.
By def\/inition, $\phi^{k, t}_n(x)$ is a~monic polynomial of degree $n$.
If some constant $h^{k, t}_0 \in \mathbb R\setminus\{0\}$ is f\/ixed, then Favard's theorem~\cite{chihara1978iop}
provides a~unique linear functional $\mathcal L^{k, t}\colon \mathbb R[x] \to \mathbb R$ such that the
orthogonality relation
\begin{gather}
\label{eq:orthogonality-op}
\mathcal L^{k,t}[x^m\phi^{k,t}_n(x)]=h^{k,t}_n\delta_{m,n},
\qquad
n=0,1,2,\dots,
\qquad
m=0,1,\dots,n,
\end{gather}
holds, where
\begin{gather*}
h^{k,t}_n=h^{k,t}_0b^{k,t}_1b^{k,t}_2\cdots b^{k,t}_n,
\qquad
n=1,2,3,\dots,
\end{gather*}
and $\delta_{m, n}$ is the Kronecker delta.

\begin{figure}[htbp]
 \centering \includegraphics{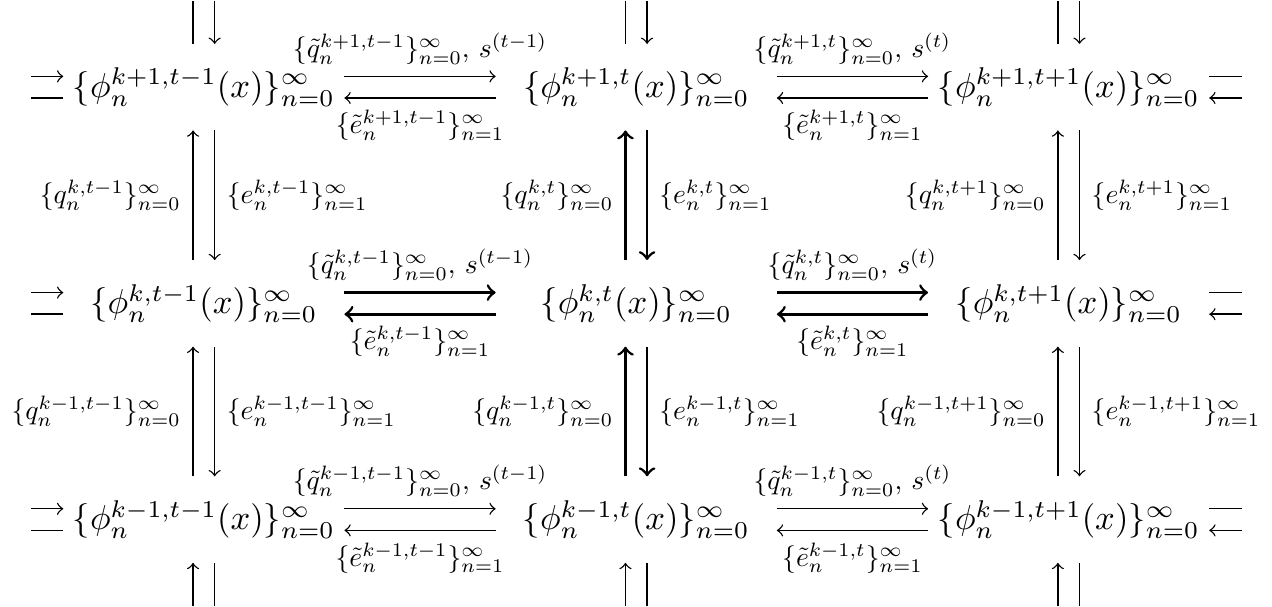}
\caption{Chain of the spectral transformations for monic orthogonal polynomials.}
\label{fig:st}
\end{figure}

Let us introduce time evolution into the orthogonal polynomials through spectral transformations.
First, the spectral transformations for the $k$-direction are
\begin{subequations}
\label{eq:st-op-k}
\begin{gather}
x\phi^{k+1,t}_n(x)=\phi^{k,t}_{n+1}(x)+q^{k,t}_{n}\phi^{k,t}_n(x),
\\
\phi^{k,t}_n(x)=\phi^{k+1,t}_n(x)+e^{k,t}_{n}\phi^{k+1,t}_{n-1}(x),
\end{gather}
\end{subequations}
where
\begin{gather}
q^{k,t}_{n}\coloneqq-\frac{\phi^{k,t}_{n+1}(0)}{\phi^{k,t}_n(0)},
\qquad
e^{k,t}_{n}\coloneqq\frac{\mathcal L^{k,t}[x^n\phi^{k,t}_n(x)]}{\mathcal L^{k+1,t}[x^{n-1}\phi^{k+1,t}_{n-1}
(x)]},
\label{eq:toda-qe}
\\
\mathcal L^{k+1,t}[\pi(x)]\coloneqq\mathcal L^{k,t}[x\pi(x)]
\qquad
\text{for all~$\pi(x)\in\mathbb R[x]$}.
\label{eq:evolution-lf-op-k}
\end{gather}
It is readily verif\/ied that $\big\{\phi^{k+1, t}_n(x)\big\}_{n=0}^\infty$ are monic orthogonal polynomials with
respect to the linear functional $\mathcal L^{k+1, t}$.
Similarly, the spectral transformations for the $t$-direction are
\begin{subequations}
\label{eq:st-op-t}
\begin{gather}
\big(x+s^{(t)}\big)\phi^{k,t+1}_n(x)=\phi^{k,t}_{n+1}(x)+\tilde q^{k,t}_{n}\phi^{k,t}_n(x),
\\
\phi^{k,t}_n(x)=\phi^{k,t+1}_n(x)+\tilde e^{k,t}_{n}\phi^{k,t+1}_{n-1}(x),
\end{gather}
\end{subequations}
where $s^{(t)}$ is a~nonzero parameter depending on $t$ and
\begin{gather}
\tilde q^{k,t}_{n}\coloneqq-\frac{\phi^{k,t}_{n+1}(-s^{(t)})}{\phi^{k,t}_n(-s^{(t)})},
\qquad
\tilde e^{k,t}_{n}\coloneqq\frac{\mathcal L^{k,t}\big[x^n\phi^{k,t}_n(x)\big]}{\mathcal L^{k,t+1}\big[x^{n-1}
\phi^{k,t+1}_{n-1}(x)\big]},
\label{eq:toda-tqe}
\\
\mathcal L^{k,t+1}[\pi(x)]\coloneqq\mathcal L^{k,t}\big[\big(x+s^{(t)}\big)\pi(x)\big]
\qquad
\text{for all~$\pi(x)\in\mathbb R[x]$}.
\label{eq:evolution-lf-op-t}
\end{gather}
The only dif\/ference between the transformations for the $k$-direction~\eqref{eq:st-op-k} and the
$t$-direction~\eqref{eq:st-op-t} is the parameter $s^{(t)}$.
Fig.~\ref{fig:st} illustrates the relations among the monic orthogonal polynomials, the spectral
transformations and the dependent variables.

Relations~\eqref{eq:trr-op},~\eqref{eq:st-op-k} and~\eqref{eq:st-op-t} yield
\begin{gather*}
\phi^{k,t}_{n+1}(x)=\big(x-a^{k,t}_n\big)\phi^{k,t}_n(x)-b^{k,t}_n\phi^{k,t}_{n-1}(x)
\\
\phantom{\phi^{k,t}_{n+1}(x)}
=\big(x-\big(q^{k,t}_n+e^{k,t}_n\big)\big)\phi^{k,t}_n(x)-q^{k,t}_{n-1}e^{k,t}_n\phi^{k,t}_{n-1}(x)
\\
\phantom{\phi^{k,t}_{n+1}(x)}
=\big(x-\big(q^{k-1,t}_n+e^{k-1,t}_{n+1}\big)\big)\phi^{k,t}_n(x)-q^{k-1,t}_{n}e^{k-1,t}_n\phi^{k,t}_{n-1}(x)
\\
\phantom{\phi^{k,t}_{n+1}(x)}
=\big(x-\big(\tilde q^{k,t}_n+\tilde e^{k,t}_n-s^{(t)}\big)\big)\phi^{k,t}_n(x)
-\tilde q^{k,t}_{n-1}\tilde e^{k,t}_n\phi^{k,t}_{n-1}(x)
\\
\phantom{\phi^{k,t}_{n+1}(x)}
=\big(x-\big(\tilde q^{k,t-1}_n+\tilde e^{k,t-1}_{n+1}-s^{(t-1)}\big)\big)\phi^{k,t}_n(x)
-\tilde q^{k,t-1}_{n}\tilde e^{k,t-1}_n\phi^{k,t}_{n-1}(x).
\end{gather*}
Hence, for consistency, the compatibility conditions
\begin{subequations}\label{eq:toda}
\begin{gather}
a^{k,t}_n=q^{k,t}_n+e^{k,t}_n=q^{k-1,t}_n+e^{k-1,t}_{n+1}
=\tilde q^{k,t}_n+\tilde e^{k,t}_n-s^{(t)}=\tilde q^{k,t-1}_n+\tilde e^{k,t-1}_{n+1}-s^{(t-1)},
\\
b^{k,t}_n=q^{k,t}_{n-1}e^{k,t}_n=q^{k-1,t}_{n}e^{k-1,t}_n
=\tilde q^{k,t}_{n-1}\tilde e^{k,t}_n=\tilde q^{k,t-1}_{n}\tilde e^{k,t-1}_n,
\\
e^{k,t}_0=\tilde e^{k,t}_0=0
\qquad
\text{for all $k$ and $t$},
\end{gather}
\end{subequations}
must be satisf\/ied.
These are the time evolution equations of the semi-inf\/inite nd-Toda lattice.
Equations~\eqref{eq:toda} give the relations among the recurrence coef\/f\/icients of
$\big\{\phi^{k,t}_n(x)\big\}_{n=0}^\infty$ and the dependent variables around $\big\{\phi^{k, t}_n(x)\big\}_{n=0}^\infty$ in the
diagram (Fig.~\ref{fig:st}).

Def\/ine the moment of the linear functional $\mathcal L^{0, t}$ by
\begin{gather*}
\mu^{(t)}_m\coloneqq\mathcal L^{0,t}\big[x^m\big].
\end{gather*}
Note that~\eqref{eq:evolution-lf-op-k} gives the relation
\begin{gather*}
\mathcal L^{k,t}\big[x^m\big]=\mathcal L^{0,t}\big[x^{k+m}\big]=\mu^{(t)}_{k+m}.
\end{gather*}
Further,~\eqref{eq:evolution-lf-op-t} gives the dispersion relation
\begin{gather}
\label{eq:dispersion-toda}
\mu^{(t+1)}_m=\mu^{(t)}_{m+1}+s^{(t)}\mu^{(t)}_m.
\end{gather}
We should remark that, if a~concrete representation of the initial linear functional $\mathcal L^{0, 0}$ is
given by a~weighted integral, then the moment may be represented concretely as
\begin{gather*}
\mu^{(t)}_m=\int_{\Omega}w(x)x^m\prod_{t'=0}^{t-1}\big(x+s^{(t')}\big)\,\mathrm{d} x,
\end{gather*}
where $\Omega$ is some interval on the real line and $w(x)$ is a~weight function def\/ined on $\Omega$.

The determinant expression of the monic orthogonal polynomials $\big\{\phi^{k, t}_n(x)\big\}_{n=0}^\infty$ is given
by
\begin{gather}
\label{eq:det-op}
\phi^{k,t}_n(x)=\frac{1}{\tau^{k,t}_n}
\begin{vmatrix}
\mu^{(t)}_k&\mu^{(t)}_{k+1}&\cdots&\mu^{(t)}_{k+n-1}&\mu^{(t)}_{k+n}
\\[2.3mm]
\mu^{(t)}_{k+1}&\mu^{(t)}_{k+2}&\cdots&\mu^{(t)}_{k+n}&\mu^{(t)}_{k+n+1}
\\[1.3mm]
\vdots&\vdots&&\vdots&\vdots
\\[1.3mm]
\mu^{(t)}_{k+n-1}&\mu^{(t)}_{k+n}&\cdots&\mu^{(t)}_{k+2n-2}&\mu^{(t)}
_{k+2n-1}
\\[2.3mm]
1&x&\cdots&x^{n-1}&x^n
\end{vmatrix},
\end{gather}
where $\tau^{k, t}_n$ is the Hankel determinant of order $n$:
\begin{gather*}
\tau^{k,t}_{-1}\coloneqq0,
\qquad
\tau^{k,t}_{0}\coloneqq1,
\qquad
\tau^{k,t}_n\coloneqq\big|\mu^{(t)}_{k+i+j}\big|_{0\le i,j\le n-1},
\qquad
n=1,2,3,\dots.
\end{gather*}
One can readily verify that the right hand side of~\eqref{eq:det-op} is a~monic polynomial of degree $n$
and satisf\/ies the orthogonality relation~\eqref{eq:orthogonality-op}.
This determinant expression~\eqref{eq:det-op} and the dispersion relation~\eqref{eq:dispersion-toda} enable
us to give Hankel determinant solutions to the nd-Toda lattice~\eqref{eq:toda}; from~\eqref{eq:toda-qe}
and~\eqref{eq:toda-tqe}, we obtain
\begin{subequations}
\label{eq:sol-toda}
\begin{gather}
q^{k,t}_n=\frac{\tau^{k,t}_n\tau^{k+1,t}_{n+1}}{\tau^{k,t}_{n+1}\tau^{k+1,t}_n},
\qquad
e^{k,t}_n=\frac{\tau^{k,t}_{n+1}\tau^{k+1,t}_{n-1}}{\tau^{k,t}_n\tau^{k+1,t}_n},
\\
\tilde q^{k,t}_n=\frac{\tau^{k,t}_n\tau^{k,t+1}_{n+1}}{\tau^{k,t}_{n+1}\tau^{k,t+1}_n},
\qquad
\tilde e^{k,t}_n=\frac{\tau^{k,t}_{n+1}\tau^{k,t+1}_{n-1}}{\tau^{k,t}_n\tau^{k,t+1}_n}.
\end{gather}
\end{subequations}

\subsection{Symmetric orthogonal polynomials and the nd-LV lattice}

Next, we consider the polynomial sequence $\big\{\sigma^{k, t}_{n}(x)\big\}_{n=0}^\infty$ def\/ined by
\begin{gather*}
\sigma^{k,t}_{2n+i}(x)\coloneqq x^i\phi^{k+i,t}_n\big(x^2\big),
\qquad
n=0,1,2,\dots,
\qquad
i=0,1.
\end{gather*}
By def\/inition, $\sigma^{k, t}_n(x)$ is a~monic polynomial of degree $n$ and has the symmetry property
\begin{gather*}
\sigma^{k,t}_n(-x)=(-1)^n\sigma^{k,t}_n(x).
\end{gather*}
Further, $\{\sigma^{k, t}_n(x)\}_{n=0}^\infty$ are orthogonal with respect to the linear functional
$\mathcal S^{k, t}$ def\/ined by{\samepage
\begin{gather*}
\mathcal S^{k,t}\big[x^{2m}\big]\coloneqq\mathcal L^{k,t}\big[x^m\big],
\qquad
\mathcal S^{k,t}\big[x^{2m+1}\big]\coloneqq0,
\qquad
m=0,1,2,\dots.
\end{gather*}
 $\big\{\sigma^{k, t}_n(x)\big\}_{n=0}^\infty$ are called monic symmetric orthogonal polynomials.}

From~\eqref{eq:st-op-k}, we have the relations
\begin{gather*}
x^2\phi^{k+1,t}_n\big(x^2\big)=\phi^{k,t}_{n+1}\big(x^2\big)+q^{k,t}_n\phi^{k,t}_n\big(x^2\big),
\\
x\phi^{k,t}_n\big(x^2\big)=x\phi^{k+1,t}_n\big(x^2\big)+e^{k,t}_n x\phi^{k+1,t}_{n-1}\big(x^2\big).
\end{gather*}
These relations lead us to the three-term recurrence relation that $\{\sigma^{k, t}_n(x)\}_{n=0}^\infty$
satisfy:
\begin{subequations}
\label{eq:trr-sop}
\begin{gather}
\sigma^{k,t}_{2n+2}(x)=x\sigma^{k,t}_{2n+1}(x)-q^{k,t}_n\sigma^{k,t}_{2n}(x),
\\
\sigma^{k,t}_{2n+1}(x)=x\sigma^{k,t}_{2n}(x)-e^{k,t}_n\sigma^{k,t}_{2n-1}(x).
\end{gather}
\end{subequations}
Spectral transformations for $\{\sigma^{k, t}_n(x)\}_{n=0}^\infty$ are also induced from~\eqref{eq:st-op-t}:
\begin{subequations}
\label{eq:st-sop}
\begin{gather}
\big(x^2+s^{(t)}\big)\sigma^{k,t+1}_{2n}(x)=\sigma^{k,t}_{2n+2}(x)+\tilde q^{k,t}_n\sigma^{k,t}_{2n}(x),
\\
\big(x^2+s^{(t)}\big)\sigma^{k,t+1}_{2n+1}(x)=\sigma^{k,t}_{2n+3}(x)+\tilde q^{k+1,t}_n\sigma^{k,t}_{2n+1}
(x),
\\
\sigma^{k,t}_{2n}(x)=\sigma^{k,t+1}_{2n}(x)+\tilde e^{k,t}_n\sigma^{k,t+1}_{2n-2}(x),
\\
\sigma^{k,t}_{2n+1}(x)=\sigma^{k,t+1}_{2n+1}(x)+\tilde e^{k+1,t}_n\sigma^{k,t+1}_{2n-1}(x).
\end{gather}
\end{subequations}
Relations~\eqref{eq:trr-sop} and~\eqref{eq:st-sop} show that there exist variables $v^{k, t}_n$ satisfying
the relations
\begin{subequations}
\label{eq:st-sop-t}
\begin{gather}
\big(x^2+s^{(t)}\big)\sigma^{k,t+1}_n(x)=x\sigma^{k,t}_{n+1}(x)+\big(s^{(t)}+v^{k,t}_n\big)\sigma^{k,t}_n(x),
\\
\big(s^{(t)}+v^{k,t}_n\big)\sigma^{k,t}_n(x)=s^{(t)}\sigma^{k,t+1}_n(x)+v^{k,t}_n x\sigma^{k,t+1}_{n-1}(x).
\end{gather}
\end{subequations}
Relations~\eqref{eq:st-sop-t} yield
\begin{gather}
\sigma^{k,t}_{n+1}(x)=x\sigma^{k,t}_n(x)-v^{k,t}_n\big(1+\big(s^{(t)}\big)^{-1}v^{k,t}_{n-1}\big)\sigma^{k,t}_{n-1}(x)
\nonumber
\\
\phantom{\sigma^{k,t}_{n+1}(x)}
=x\sigma^{k,t}_n(x)-v^{k,t-1}_n\big(1+\big(s^{(t-1)}\big)^{-1}v^{k,t-1}_{n+1}\big)\sigma^{k,t}_{n-1}(x).\label{eq:st-compat}
\end{gather}
Hence, the compatibility condition
\begin{subequations}
\label{eq:nd-LV}
\begin{gather}
v^{k,t}_n\big(1+\big(s^{(t)}\big)^{-1}v^{k,t}_{n-1}\big)=v^{k,t-1}_n\big(1+\big(s^{(t-1)}\big)^{-1}v^{k,t-1}_{n+1}\big),
\\
v^{k,t}_0=0
\qquad
\text{for all~$k$ and $t$},
\end{gather}
\end{subequations}
must be satisf\/ied.
This is the time evolution equation of the semi-inf\/inite nd-LV lattice.

From relations~\eqref{eq:trr-sop}--\eqref{eq:st-compat}, we obtain the Miura transformation between the
nd-Toda lattice~\eqref{eq:toda} and the nd-LV lattice~\eqref{eq:nd-LV}:
\begin{gather*}
q^{k,t}_n
=v^{k,t}_{2n+1}\big(1+\big(s^{(t)}\big)^{-1}v^{k,t}_{2n}\big)=v^{k,t-1}_{2n+1}\big(1+\big(s^{(t-1)}\big)^{-1}
v^{k,t-1}_{2n+2}\big),
\\
e^{k,t}_n
=v^{k,t}_{2n}\big(1+\big(s^{(t)}\big)^{-1}v^{k,t}_{2n-1}\big)=v^{k,t-1}_{2n}\big(1+\big(s^{(t-1)}\big)^{-1}
v^{k,t-1}_{2n+1}\big),
\\
\tilde q^{k,t}_n
=s^{(t)}\big(1+\big(s^{(t)}\big)^{-1}v^{k,t}_{2n+1}\big)\big(1+\big(s^{(t)}\big)^{-1}v^{k,t}_{2n}\big)
\\
\phantom{\tilde q^{k,t}_n}
=s^{(t)}\big(1+\big(s^{(t)}\big)^{-1}v^{k-1,t}_{2n+1}\big)\big(1+\big(s^{(t)}\big)^{-1}v^{k-1,t}_{2n+2}\big),
\\
\tilde e^{k,t}_n
=\big(s^{(t)}\big)^{-1}v^{k,t}_{2n}v^{k,t}_{2n-1}=\big(s^{(t)}\big)^{-1}v^{k-1,t}_{2n}v^{k-1,t}_{2n+1}.
\end{gather*}
In addition, from relations~\eqref{eq:st-sop-t}, we obtain
\begin{gather*}
1+\big(s^{(t)}\big)^{-1}v^{k,t}_{2n}=\frac{\sigma^{k,t+1}_{2n}(0)}{\sigma^{k,t}_{2n}(0)}=\frac{\phi^{k,t+1}_n(0)}
{\phi^{k,t}_n(0)},
\\
1+\big(s^{(t)}\big)^{-1}v^{k,t}_{2n-1}=\big({-}s^{(t)}\big)^{-1/2}\frac{\sigma^{k,t}_{2n}\big((-s^{(t)})^{1/2}\big)}
{\sigma^{k,t}_{2n-1}\big((-s^{(t)})^{1/2}\big)}=-\big(s^{(t)}\big)^{-1}\frac{\phi^{k,t}_n(-s^{(t)})}{\phi^{k+1,t}
_{n-1}(-s^{(t)})}.
\end{gather*}
By using these relations and the solutions to the nd-Toda lattice~\eqref{eq:sol-toda}, we obtain Hankel
determinant solutions to the nd-LV lattice~\eqref{eq:nd-LV}:
\begin{gather*}
v^{k,t}_{2n+1}=q^{k,t}_n\frac{\phi^{k,t}_n(0)}{\phi^{k,t+1}_n(0)}=\frac{\tau^{k,t+1}_n\tau^{k+1,t}_{n+1}}
{\tau^{k,t}_{n+1}\tau^{k+1,t+1}_n},
\\
v^{k,t}_{2n}=-s^{(t)}e^{k,t}_n\frac{\phi^{k+1,t}_{n-1}(-s^{(t)})}{\phi^{k,t}_n(-s^{(t)})}=s^{(t)}
\frac{\tau^{k,t}_{n+1}\tau^{k+1,t+1}_{n-1}}{\tau^{k,t+1}_n\tau^{k+1,t}_n}.
\end{gather*}

\section{Derivation of the $\text{R}_{\text{II}}$ chain and the nd-mKdV lattice}
\label{sec:derivation-rii-chain}

We will apply the framework constructed in the previous section to monic $\text{R}_{\text{II}}$ polynomials and derive the
nd-mKdV lattice.

\subsection{$\text{R}_{\text{II}}$ polynomials and the $\text{R}_{\text{II}}$ chain}

Monic $\text{R}_{\text{II}}$ polynomials are def\/ined by the three-term recurrence relation of the form
\begin{gather*}
\varphi^{k,t}_{-1}(x)\coloneqq0,
\qquad
\varphi^{k,t}_0(x)\coloneqq1,
\\
\varphi^{k,t}_{n+1}(x)\coloneqq\big(\big(1+\beta^{k,t}_n\big)x-\alpha^{k,t}_n\big)\varphi^{k,t}_n(x)-\beta^{k,t}
_n(x+\gamma_{k+t+2n-2})(x+\gamma_{k+t+2n-1})\varphi^{k,t}_{n-1}(x),
\\
n=0,1,2,\dots,
\end{gather*}
where $\alpha^{k, t}_n \in \mathbb R$ and $\beta^{k, t}_n, \gamma_{k+t+n} \in \mathbb R\setminus\{0\}$.
If nonzero constants $h^{k, t}_0$ and $h^{k, t}_1$ are f\/ixed, then a~Favard-type
theorem~\cite{ismail1995goa} guarantees the existence of a~unique linear functional $\mathcal L^{k, t}$
such that the orthogonality relation
\begin{gather*}
\mathcal L^{k,t}\left[\frac{x^m\varphi^{k,t}_n(x)}{\prod_{j=0}^{2n-1}(x+\gamma_{k+t+j})}\right]=h^{k,t}
_n\delta_{m,n},
\qquad
n=0,1,2,\dots,
\qquad
m=0,1,\dots,n,
\end{gather*}
holds, where $h^{k, t}_n$, $n=2, 3, \dots$ are nonzero constants.
Note that $\mathcal L^{k, t}$ is def\/ined on the vector space spanned by
$\frac{1}{\prod_{j=0}^{l-1}(x+\gamma_{k+t+j})}$, $l=0, 1, 2, \dots$.

As in the case of monic orthogonal polynomials, we introduce the time evolution of the monic $\text{R}_{\text{II}}$
polynomials by the following spectral transformations:
\begin{subequations}
\label{eq:st-rii}
\begin{gather}
\big(1+q^{k,t}_n\big)x\varphi^{k+1,t}_n(x)=\varphi^{k,t}_{n+1}(x)+q^{k,t}_n(x+\gamma_{k+t+2n})\varphi^{k,t}
_n(x),
\\
\big(1+e^{k,t}_n\big)\varphi^{k,t}_n(x)=\varphi^{k+1,t}_n(x)+e^{k,t}_n(x+\gamma_{k+t+2n-1})\varphi^{k+1,t}
_{n-1}(x),
\\
\big(1+\tilde q^{k,t}_n\big)\big(x+s^{(t)}\big)\varphi^{k,t+1}_n(x)=\varphi^{k,t}_{n+1}(x)+\tilde q^{k,t}
_n(x+\gamma_{k+t+2n})\varphi^{k,t}_n(x),
\\
\big(1+\tilde e^{k,t}_n\big)\varphi^{k,t}_n(x)=\varphi^{k,t+1}_n(x)+\tilde e^{k,t}_n(x+\gamma_{k+t+2n-1}
)\varphi^{k,t+1}_{n-1}(x).
\end{gather}
\end{subequations}
These spectral transformations were originally introduced by Zhedanov~\cite{zhedanov1999brf}.
By choosing the variables $q^{k, t}_n$, $e^{k, t}_n$, $\tilde q^{k, t}_n$ and $\tilde e^{k, t}_n$ as above,
the leading coef\/f\/icients of both sides of~\eqref{eq:st-rii} become equal.
The time evolution of the linear functional is also given by
\begin{gather}
\label{eq:evolution-lf-rii}
\mathcal L^{k+1,t}[\rho(x)]\coloneqq\mathcal L^{k,t}\left[\frac{x}{x+\gamma_{k+t}}\rho(x)\right],
\qquad
\mathcal L^{k,t+1}[\rho(x)]\coloneqq\mathcal L^{k,t}\left[\frac{x+s^{(t)}}{x+\gamma_{k+t}}\rho(x)\right]
\end{gather}
for rational functions $\rho(x)$.
One can verify that $\big\{\varphi^{k+1, t}_n(x)\big\}_{n=0}^\infty$ and $\big\{\varphi^{k, t+1}_n(x)\big\}_{n=0}^\infty$
are both also monic $\text{R}_{\text{II}}$ polynomials.
The spectral transformations~\eqref{eq:st-rii} induce the time evolution equations of the semi-inf\/inite
monic-type $\text{R}_{\text{II}}$ chain:
\begin{subequations}
\label{eq:rii}
\begin{gather}
\alpha^{k,t}_n=\gamma_{k+t+2n}q^{k,t}_n+\gamma_{k+t+2n-1}e^{k,t}_n\frac{1+q^{k,t}_n}{1+q^{k,t}_{n-1}}
\nonumber
\\
\phantom{\alpha^{k,t}_n}
=\gamma_{k+t+2n-1}q^{k-1,t}_n\frac{1+e^{k-1,t}_{n+1}}{1+e^{k-1,t}_n}+\gamma_{k+t+2n}e^{k-1,t}_{n+1}
\nonumber
\\
\phantom{\alpha^{k,t}_n}
=\gamma_{k+t+2n}\tilde q^{k,t}_n+\gamma_{k+t+2n-1}\tilde e^{k,t}_n\frac{1+\tilde q^{k,t}_n}
{1+\tilde q^{k,t}_{n-1}}-s^{(t)}\big(1+\tilde q^{k,t}_n\big)\big(1+\tilde e^{k,t}_n\big)
\nonumber
\\
\phantom{\alpha^{k,t}_n}
=\gamma_{k+t+2n-1}\tilde q^{k,t-1}_n\frac{1+\tilde e^{k,t-1}_{n+1}}{1+\tilde e^{k,t-1}_n}+\gamma_{k+t+2n}
\tilde e^{k,t-1}_{n+1}-s^{(t-1)}\big(1+\tilde q^{k,t-1}_n\big)\big(1+\tilde e^{k,t-1}_{n+1}\big),
\\
\beta^{k,t}_n=q^{k,t}_{n-1}e^{k,t}_n\frac{1+q^{k,t}_n}{1+q^{k,t}_{n-1}}=q^{k-1,t}_n e^{k-1,t}
_n\frac{1+e^{k-1,t}_{n+1}}{1+e^{k-1,t}_n}
\nonumber
\\
\phantom{\beta^{k,t}_n}
=\tilde q^{k,t}_{n-1}\tilde e^{k,t}_n\frac{1+\tilde q^{k,t}_n}{1+\tilde q^{k,t}_{n-1}}=\tilde q^{k,t-1}
_n\tilde e^{k,t-1}_n\frac{1+\tilde e^{k,t-1}_{n+1}}{1+\tilde e^{k,t-1}_n},
\\
e^{k,t}_0=\tilde e^{k,t}_0=0
\qquad
\text{for all~$k$ and $t$}.
\end{gather}
\end{subequations}
Note that the $\text{R}_{\text{II}}$ chain was originally introduced by Spiridonov and Zhedanov~\cite{spiridonov2000stc}.
The original chain is described by three equations and four types of dependent variables with one
constraint.
The monic-type version~\eqref{eq:rii} is, however, described by essentially only two equations and two
types of dependent variables; since we are now considering two time variables $k$ and $t$, there are four
types of dependent variables.

We def\/ine the moment of the linear functional $\mathcal L^{0, t}$ by
\begin{gather*}
\mu^{(t)}_{m,l}\coloneqq\mathcal L^{0,t}\left[\frac{x^m}{\prod_{j=0}^{l-1}(x+\gamma_{t+j})}\right].
\end{gather*}
Note that the time evolution of the linear functional~\eqref{eq:evolution-lf-rii} gives the relation
\begin{gather*}
\mathcal L^{k,t}\left[\frac{x^m}{\prod_{j=0}^{l-1}(x+\gamma_{k+t+j})}\right]=\mathcal L^{0,t}
\left[\frac{x^{k+m}}{\prod_{j=0}^{k+l-1}(x+\gamma_{t+j})}\right]=\mu^{(t)}_{k+m,k+l}
\end{gather*}
and the dispersion relations
\begin{gather}
\label{eq:dispersion-rii}
\mu^{(t+1)}_{m,l}=\mu^{(t)}_{m+1,l+1}+s^{(t)}\mu^{(t)}_{m,l+1},
\qquad
\mu^{(t)}_{m,l}=\mu^{(t)}_{m+1,l+1}+\gamma_{t+l}\mu^{(t)}_{m,l+1}.
\end{gather}
Then, the determinant expression of the monic $\text{R}_{\text{II}}$ polynomials $\big\{\varphi^{k, t}_n(x)\big\}_{n=0}^\infty$ is
given by
\begin{gather}
\label{eq:det-rii}
\varphi^{k,t}_n(x)=\frac{1}{\tau^{k,2n,t}_{n}}
\begin{vmatrix}
\mu^{(t)}_{k,k+2n}&\mu^{(t)}_{k+1,k+2n}&\cdots&\mu^{(t)}_{k+n-1,k+2n}&\mu^{(t)}_{k+n,k+2n}
\\[2.3mm]
\mu^{(t)}_{k+1,k+2n}&\mu^{(t)}_{k+2,k+2n}&\cdots&\mu^{(t)}_{k+n,k+2n}&\mu^{(t)}_{k+n+1,k+2n}
\\[1.3mm]
\vdots&\vdots&&\vdots&\vdots
\\[1.3mm]
\mu^{(t)}_{k+n-1,k+2n}&\mu^{(t)}_{k+n,k+2n}&\cdots&\mu^{(t)}_{k+2n-2,k+2n}
&\mu^{(t)}_{k+2n-1,k+2n}
\\[2.3mm]
1&x&\cdots&x^{n-1}&x^n
\end{vmatrix}
,
\end{gather}
where $\tau^{k, l, t}_n$ is the Hankel determinant of order $n$:
\begin{gather}
\label{eq:hankel-rii}
\tau^{k,l,t}_{-1}\coloneqq0,
\qquad
\tau^{k,l,t}_{0}\coloneqq1,
\qquad
\tau^{k,l,t}_n\coloneqq\big|\mu^{(t)}_{k+i+j,k+l}\big|_{0\le i,j\le n-1},
\qquad
n=1,2,3,\dots.
\end{gather}
We should remark that the Casorati determinant representation of the $\text{R}_{\text{II}}$ polynomials was found by
Spiridonov and Zhedanov~\cite{spiridonov2003tbr}.
The Hankel determinant expression given above ref\/lects the structure of the discrete two-dimensional Toda
hierarchy~\cite{tsujimoto2010dso}.
By using the determinant expression~\eqref{eq:det-rii}, the dispersion relation~\eqref{eq:dispersion-rii},
and a~determinant identity called Pl\"ucker relation, we can f\/ind the Hankel determinant solutions to the
semi-inf\/inite monic-type $\text{R}_{\text{II}}$ chain:
\begin{gather*}
q^{k, t}_n=(\gamma_{k+t+2n})^{-1}\frac{\tau^{k, 2n, t}_n\tau^{k+1, 2n+1, t}_{n+1}}{\tau^{k, 2n+2,
t}_{n+1}\tau^{k+1, 2n-1, t}_n},
\qquad
e^{k, t}_n=\gamma_{k+t+2n}\frac{\tau^{k, 2n+1, t}_{n+1}\tau^{k+1, 2n-2, t}_{n-1}}{\tau^{k, 2n-1,
t}_n\tau^{k+1, 2n, t}_n},
\\
\tilde q^{k, t}_n=\big(\gamma_{k+t+2n}-s^{(t)}\big)^{-1}\frac{\tau^{k, 2n, t}_n \tau^{k, 2n+1,
t+1}_{n+1}}{\tau^{k, 2n+2, t}_{n+1}\tau^{k, 2n-1, t+1}_n},
\quad \
\tilde e^{k, t}_n=\big(\gamma_{k+t+2n}-s^{(t)}\big)\frac{\tau^{k, 2n+1, t}_{n+1}\tau^{k, 2n-2,
t+1}_{n-1}}{\tau^{k, 2n-1, t}_n\tau^{k, 2n, t+1}_n}.\!
\end{gather*}

\subsection{Symmetric $\text{R}_{\text{II}}$ polynomials and the nd-mKdV lattice}

We introduce a~symmetric version of the monic $\text{R}_{\text{II}}$ polynomials, which is an analogue of the monic symmetric
orthogonal polynomials.

Let us def\/ine a~polynomial sequence $\big\{\varsigma^{k, t}_n(x)\big\}_{n=0}^\infty$ by
\begin{gather*}
\varsigma^{k,t}_{2n+i}(x)\coloneqq x^i\varphi^{k+i,t}_n\big(x^2\big),
\qquad
n=0,1,2,\dots,
\qquad
i=0,1.
\end{gather*}
The corresponding linear functional $\mathcal S^{k, t}$ is given by
\begin{gather*}
\mathcal S^{k,t}\left[\frac{x^{2m}}{\prod_{j=0}^{l-1}(x^2+\gamma_{k+t+j})}\right]\coloneqq\mathcal L^{k,t}
\left[\frac{x^{m}}{\prod_{j=0}^{l-1}(x+\gamma_{k+t+j})}\right],
\\
\mathcal S^{k,t}\left[\frac{x^{2m+1}}
{\prod_{j=0}^{l-1}(x^2+\gamma_{k+t+j})}\right]\coloneqq0,
\qquad
l=0,1,2,\dots,
\qquad
m=0,1,\dots,l.
\end{gather*}
The spectral transformations~\eqref{eq:st-rii} yield the three-term recurrence relations
\begin{subequations}
\label{eq:trr-srii}
\begin{gather}
\varsigma^{k,t}_{2n+2}(x)=\big(1+q^{k,t}_n\big)x\varsigma^{k,t}_{2n+1}(x)-q^{k,t}_n\big(x^2+\gamma_{k+t+2n}
\big)\varsigma^{k,t}_{2n}(x),
\\
\varsigma^{k,t}_{2n+1}(x)=\big(1+e^{k,t}_n\big)x\varsigma^{k,t}_{2n}(x)-e^{k,t}_n\big(x^2+\gamma_{k+t+2n-1}
\big)\varsigma^{k,t}_{2n-1}(x),
\end{gather}
\end{subequations}
and the spectral transformations for the monic symmetric $\text{R}_{\text{II}}$ polynomials $\big\{\varsigma^{k,
t}_n(x)\big\}_{n=0}^\infty$:
\begin{subequations}
\label{eq:st-srii}
\begin{gather}
\big(1+\tilde q^{k,t}_n\big)\big(x^2+s^{(t)}\big)\varsigma^{k,t+1}_{2n}(x)
=\varsigma^{k,t}_{2n+2}(x)+\tilde q^{k,t}_n\big(x^2+\gamma_{k+t+2n}\big)\varsigma^{k,t}_{2n}(x),
\\
\big(1+\tilde q^{k+1,t}_n\big)\big(x^2+s^{(t)}\big)\varsigma^{k,t+1}_{2n+1}(x)=\varsigma^{k,t}_{2n+3}
(x)+\tilde q^{k+1,t}_n\big(x^2+\gamma_{k+t+2n+1}\big)\varsigma^{k,t}_{2n+1}(x),
\\
\big(1+\tilde e^{k,t}_n\big)\varsigma^{k,t}_{2n}(x)=\varsigma^{k,t+1}_{2n}(x)+\tilde e^{k,t}
_n\big(x^2+\gamma_{k+t+2n-1}\big)\varsigma^{k,t+1}_{2n-2}(x),
\\
\big(1+\tilde e^{k+1,t}_n\big)\varsigma^{k,t}_{2n+1}(x)=\varsigma^{k,t+1}_{2n+1}(x)+\tilde e^{k+1,t}
_n\big(x^2+\gamma_{k+t+2n}\big)\varsigma^{k,t+1}_{2n-1}(x).
\end{gather}
\end{subequations}
Relations~\eqref{eq:trr-srii} and~\eqref{eq:st-srii} show that there exist variables $v^{k, t}_n$
satisfying the relations
\begin{subequations}
\label{eq:st-srii-t}
\begin{gather}
\big(\gamma_{k+t+n}+v^{k,t}_n\big)\big(x^2+s^{(t)}\big)\varsigma^{k,t+1}_n(x)
\nonumber
\\
\qquad {}=\big(\gamma_{k+t+n}-s^{(t)}\big)x\varsigma^{k,t}_{n+1}(x)
+\big(s^{(t)}+v^{k,t}_n\big)\big(x^2+\gamma_{k+t+n}\big)\varsigma^{k,t}_n(x),
\\
\gamma_{k+t+n}\big(s^{(t)}+v^{k,t}_n\big)\varsigma^{k,t}_n(x)
\nonumber
\\
\qquad{}
=s^{(t)}(\gamma_{k+t+n}+v^{k,t}_n)\varsigma^{k,t+1}_n(x)
+\big(\gamma_{k+t+n}-s^{(t)}\big)v^{k,t}_n x\varsigma^{k,t+1}_{n-1}(x).
\end{gather}
\end{subequations}
Relations~\eqref{eq:st-srii-t} yield
\begin{gather*}
\varsigma^{k,t}_{n+1}(x)=
\left(1+(\gamma_{k+t+n-1})^{-1}v^{k,t}_n\frac{1+\big(s^{(t)}\big)^{-1}v^{k,t}_{n-1}}{1+(\gamma_{k+t+n-1})^{-1}
v^{k,t}_{n-1}}\right)x\varsigma^{k,t}_n(x)
\\
\phantom{\varsigma^{k,t}_{n+1}(x)=}
{}-(\gamma_{k+t+n-1})^{-1}v^{k,t}_n\frac{1+\big(s^{(t)}\big)^{-1}v^{k,t}_{n-1}}{1+(\gamma_{k+t+n-1})^{-1}v^{k,t}_{n-1}
}\big(x^2+\gamma_{k+t+n-1}\big)\varsigma^{k,t}_{n-1}(x)
\\
\phantom{\varsigma^{k,t}_{n+1}(x)}{}
=\left(1+(\gamma_{k+t+n-1})^{-1}v^{k,t-1}_n\frac{1+\big(s^{(t-1)}\big)^{-1}v^{k,t-1}_{n+1}}{1+(\gamma_{k+t+n}
)^{-1}v^{k,t-1}_{n+1}}\right)x\varsigma^{k,t}_n(x)
\\
\phantom{\varsigma^{k,t}_{n+1}(x)=}
{}-(\gamma_{k+t+n-1})^{-1}v^{k,t-1}_n\frac{1+\big(s^{(t-1)}\big)^{-1}v^{k,t-1}_{n+1}}{1+(\gamma_{k+t+n})^{-1}v^{k,t-1}
_{n+1}}\big(x^2+\gamma_{k+t+n-1}\big)\varsigma^{k,t}_{n-1}(x).
\end{gather*}
Hence, the compatibility condition
\begin{subequations}
\label{eq:nd-mKdV}
\begin{gather}
v^{k,t}_n\frac{1+\big(s^{(t)}\big)^{-1}v^{k,t}_{n-1}}{1+(\gamma_{k+t+n-1})^{-1}v^{k,t}_{n-1}}=v^{k,t-1}
_n\frac{1+\big(s^{(t-1)}\big)^{-1}v^{k,t-1}_{n+1}}{1+(\gamma_{k+t+n})^{-1}v^{k,t-1}_{n+1}},
\\
v^{k,t}_0=0
\qquad
\text{for all~$k$ and $t$},
\end{gather}
\end{subequations}
must be satisf\/ied.
This is the time evolution equation of the semi-inf\/inite nd-mKdV lattice, a~nonautonomous version of the
discrete mKdV lattice~\cite{takahashi1997bbs}.
Note that the nd-mKdV lattice~\eqref{eq:nd-mKdV} reduces to the nd-LV lattice~\eqref{eq:nd-LV} as
$\gamma_{k+t+n} \to \infty$.

The Miura transformation between the monic-type $\text{R}_{\text{II}}$ chain~\eqref{eq:rii} and the nd-mKdV
lattice~\eqref{eq:nd-mKdV} is obtained as follows:
\begin{gather*}
q^{k,t}_n=(\gamma_{k+t+2n})^{-1}v^{k,t}_{2n+1}
\frac{1+\big(s^{(t)}\big)^{-1}v^{k,t}_{2n}}{1+(\gamma_{k+t+2n})^{-1}
v^{k,t}_{2n}}
=(\gamma_{k+t+2n})^{-1}v^{k,t-1}_{2n+1}
\frac{1+\big(s^{(t-1)}\big)^{-1}v^{k,t-1}_{2n+2}}{1+(\gamma_{k+t+2n+1})^{-1}
v^{k,t-1}_{2n+2}},
\\
e^{k,t}_n=(\gamma_{k+t+2n-1})^{-1}v^{k,t}_{2n}
\frac{1+\big(s^{(t)}\big)^{-1}v^{k,t}_{2n-1}}{1+(\gamma_{k+t+2n-1})^{-1}v^{k,t}_{2n-1}}
\\
\phantom{e^{k,t}_n}{}
=(\gamma_{k+t+2n-1})^{-1}v^{k,t-1}_{2n}
\frac{1+\big(s^{(t-1)}\big)^{-1}v^{k,t-1}_{2n+1}}{1+(\gamma_{k+t+2n})^{-1}v^{k,t-1}_{2n+1}},
\\
\tilde q^{k,t}_n=\big(\gamma_{k+t+2n}-s^{(t)}\big)^{-1}s^{(t)}\big(1+\big(s^{(t)}\big)^{-1}v^{k,t}_{2n+1}
\big)\big(1+\big(s^{(t)}\big)^{-1}v^{k,t}_{2n}\big)
\\
\phantom{\tilde q^{k,t}_n}{}
=\big(\gamma_{k+t+2n}-s^{(t)}\big)^{-1}s^{(t)}\big(1+\big(s^{(t)}\big)^{-1}v^{k-1,t}_{2n+1}\big)
\big(1+\big(s^{(t)}\big)^{-1}v^{k-1,t}_{2n+2}\big),
\\
\tilde e^{k,t}_n=\frac{\big(\gamma_{k+t+2n}-s^{(t)}\big)v^{k,t}_{2n}v^{k,t}_{2n-1}}{s^{(t)}\gamma_{k+t+2n-1}
\gamma_{k+t+2n}\big(1+(\gamma_{k+t+2n})^{-1}v^{k,t}_{2n}\big)\big(1+(\gamma_{k+t+2n-1})^{-1}v^{k,t}_{2n-1}\big)}
\\
\phantom{\tilde e^{k,t}_n}{}
=\frac{\big(\gamma_{k+t+2n}-s^{(t)}\big)v^{k-1,t}_{2n}v^{k-1,t}_{2n+1}}{s^{(t)}\gamma_{k+t+2n-1}\gamma_{k+t+2n}
\big(1+(\gamma_{k+t+2n-1})^{-1}v^{k-1,t}_{2n}\big)\big(1+(\gamma_{k+t+2n})^{-1}v^{k-1,t}_{2n+1}\big)}.
\end{gather*}
Furthermore, from relations~\eqref{eq:st-srii-t}, we obtain
\begin{gather*}
\frac{1+\big(s^{(t)}\big)^{-1}v^{k,t}_{2n}}{1+(\gamma_{k+t+2n})^{-1}v^{k,t}_{2n}}=\frac{\varphi^{k,t+1}_{n}(0)}
{\varphi^{k,t}_{n}(0)},
\\
1+\big(s^{(t)}\big)^{-1}v^{k,t}_{2n-1}=(-s^{(t)})^{-1}\frac{\varphi^{k,t}_n(-s^{(t)})}{\varphi^{k+1,t}_{n-1}
(-s^{(t)})},
\\
1+(\gamma_{k+t+2n-1})^{-1}v^{k,t}_{2n-1}=(-\gamma_{k+t+2n-1})^{-1}\frac{\varphi^{k,t}_{n}(-\gamma_{k+t+2n-1}
)}{\varphi^{k+1,t+1}_{n-1}(-\gamma_{k+t+2n-1})}.
\end{gather*}
Hence, Hankel determinant solutions to the nd-mKdV lattice~\eqref{eq:nd-mKdV} are given by
\begin{gather*}
v^{k,t}_{2n+1}=\gamma_{k+t+2n}q^{k,t}_n\frac{\varphi^{k,t}_{n}(0)}{\varphi^{k,t+1}_{n}(0)}
=\frac{\tau^{k,2n,t+1}_n\tau^{k+1,2n+1,t}_{n+1}}{\tau^{k,2n+2,t}_{n+1}\tau^{k+1,2n-1,t+1}_n},
\\
v^{k,t}_{2n}=s^{(t)}e^{k,t}_n\frac{\varphi^{k+1,t}_{n-1}(-s^{(t)})}{\varphi^{k,t}_n(-s^{(t)})}
\frac{\varphi^{k,t}_{n}(-\gamma_{k+t+2n-1})}{\varphi^{k+1,t+1}_{n-1}(-\gamma_{k+t+2n-1})}=s^{(t)}
\gamma_{k+t+2n}\frac{\tau^{k,2n+1,t}_{n+1}\tau^{k+1,2n-2,t+1}_{n-1}}{\tau^{k,2n-1,t+1}_n\tau^{k+1,2n,t}_n}.
\end{gather*}

\begin{remark}
Spiridonov~\cite{spiridonov2000scp} f\/irst considered spectral transformations for the (not monic)
symmetric $\text{R}_{\text{II}}$ polynomials.
By using spectral transformations, he derived a~generalization of the nd-LV lattice~\eqref{eq:nd-LV}, which
is more complicated than the nd-mKdV lattice~\eqref{eq:nd-mKdV}.
We have considered the monic symmetric $\text{R}_{\text{II}}$ polynomials and their spectral transformations which possess
the following symmetry.
Consider an independent variable transformation $t'=-k-t-n$
and introduce $\tilde \varsigma^{k,t'}_n(x)\coloneqq \varsigma^{k, -k-t'-n+1}_n(x)$,
$\tilde v^{k, t'}_n\coloneqq v^{k, -k-t'-n}_n$,
$\tilde s_{k+t'+n}\coloneqq s^{(-k-t'-n)}$ and $\tilde \gamma^{(t')}\coloneqq \gamma_{-t'}$.
Then, the spectral transformations for the monic symmetric $\text{R}_{\text{II}}$ polynomials~\eqref{eq:st-srii-t} may be
rewritten as
\begin{gather*}
\big(\tilde s_{k+t'+n}+\tilde v^{k,t'}_n\big)\big(x^2+\tilde\gamma^{(t')}\big)\tilde\varsigma^{k,t'+1}_n(x)
\\
\qquad{}
=\big(\tilde s_{k+t'+n}-\tilde\gamma^{(t')}\big)x\tilde\varsigma^{k,t'}_{n+1}(x)
+\big(\tilde\gamma^{(t')}+\tilde v^{k,t'}_n\big)\big(x^2+\tilde s_{k+t'+n}\big)\tilde\varsigma^{k,t'}_n(x),
\\
\tilde s_{k+t'+n}\big(\tilde\gamma^{(t')}+\tilde v^{k,t'}_n\big)\tilde\varsigma^{k,t'}_n(x)
\\
\qquad{}
=\tilde\gamma^{(t')}\big(\tilde s_{k+t'+n}+\tilde v^{k,t'}_n\big)\tilde\varsigma^{k,t'+1}_n(x)
+\big(\tilde s_{k+t'+n}-\tilde\gamma^{(t')}\big)\tilde v^{k,t'}_n x\tilde\varsigma^{k,t'+1}_{n-1}(x),
\end{gather*}
so that the roles of the parameters are replaced.
Using the symmetric form of the spectral transformations~\eqref{eq:st-srii-t}, we can derive the
corresponding discrete integrable system in a~simpler form.

In another study, Spiridonov et al.~\cite{spiridonov2007idt} derived a~discrete integrable system
called the FST chain and discussed its connection to the $\text{R}_{\text{II}}$ chain.
The time evolution equation of the FST chain is
\begin{gather*}
\frac{\gamma_{k+t+n}-s^{(t)}+A^{k,t}_n A^{k,t}_{n-1}}{A^{k,t}_n}=\frac{\gamma_{k+t+n-1}-s^{(t-1)}+A^{k,t-1}
_n A^{k,t-1}_{n+1}}{A^{k,t-1}_n},
\\
A^{k,t}_{-1}=0
\qquad
\text{for all~$k$ and $t$}.
\end{gather*}
Particular solutions to the FST chain may also be expressed by the Hankel determinant~\eqref{eq:hankel-rii}:
\begin{gather*}
A^{k,t}_{2n}=\big(\gamma_{k+t+2n}-s^{(t)}\big)\frac{\tau^{k,2n+1,t}_{n+1}\tau^{k,2n-1,t+1}_n}{\tau^{k,2n,t}
_n\tau^{k,2n,t+1}_n},
\\
A^{k,t}_{2n+1}=\big(\gamma_{k+t+2n+1}-s^{(t)}\big)\frac{\tau^{k,2n+2,t}_{n+1}\tau^{k,2n,t+1}_n}{\tau^{k,2n+1,t}
_{n+1}\tau^{k,2n+1,t+1}_{n+1}}.
\end{gather*}
Similarly, we have the discrete potential KdV lattice
\begin{gather*}
\big(\epsilon^{k,t-1}_n-\epsilon^{k,t}_n\big)\big(\epsilon^{k,t-1}_{n+1}-\epsilon^{k,t}_{n-1}\big)
=\gamma_{k+t+n-1}-s^{(t-1)},
\\
\epsilon^{k,t}_{-1}=0
\qquad
\text{for all~$k$ and $t$},
\end{gather*}
and its Hankel determinant solutions
\begin{gather*}
\epsilon^{k,t}_{2n}=\frac{\tau^{k,2n,t}_{n+1}}{\tau^{k,2n,t}_n},
\qquad
\epsilon^{k,t}_{2n+1}=\frac{\tau^{k,2n+1,t}_n}{\tau^{k,2n+1,t}_{n+1}}.
\end{gather*}
Therefore, these systems and the nd-mKdV lattice~\eqref{eq:nd-mKdV} are connected via the bilinear
formalism.
\end{remark}

\section{Concluding remarks}
\label{sec:concluding-remarks}
In this paper, we developed the spectral transformation technique for symmetric $\text{R}_{\text{II}}$ polynomials and
derived the nd-mKdV lattice as the compatibility condition.
Moreover, we obtained a~direct connection between the $\text{R}_{\text{II}}$ chain and the nd-mKdV lattice.
It is easily verif\/ied by numerical experiments that the obtained nd-mKdV lattice with a~non-periodic
f\/inite lattice condition can compute the generalized eigenvalues of the tridiagonal matrix pencil that
corresponds to the $\text{R}_{\text{II}}$ polynomials through the Miura transformation.
More practical applications of the nd-mKdV lattice to numerical algorithms are left for future work.
In particular, the application to generalized singular value decomposition~\cite{vanloan1976gsv} will be
discussed in detail.

In recent studies, various discrete Painlev\'e equations have been obtained as reductions of discrete
integrable
systems~\cite{grammaticos2005ril,hay2007lpl,ormerod2012rlm,ormerod2013sss,ormerod2013tri,ormerod2013dpe}.
On the other hand, it is known that the $\text{R}_{\text{II}}$ chain and the elliptic Painlev\'e equation~\cite{sakai2001rsa}
have solutions expressible in terms of the elliptic hypergeometric function
$_{10}E_9$~\cite{kajiwara2003sep, spiridonov2000stc}.
In addition, it was pointed out that the contiguity relations of the elliptic Painlev\'e equation are
similar to the linear relations of the $\text{R}_{\text{II}}$ chain~\cite{noumi2013pie}.
Supported by these evidences, one may believe that a~reduction of the $\text{R}_{\text{II}}$ chain may give rise to the
elliptic Painlev\'e equation.
This work linked the nd-mKdV lattice with the $\text{R}_{\text{II}}$ chain.
We are now concerned with its relationship to the discrete Painlev\'e equations.
In particular, we expect that the elliptic Painlev\'e equation will appear as a~reduction of the nd-mKdV
lattice.

\subsection*{Acknowledgements}

The authors thank Professor Kenji Kajiwara for fruitful comments and the
anonymous referees for their valuable suggestions.
This work was supported by JSPS KAKENHI Grant Numbers 11J04105 and 25400110.

\pdfbookmark[1]{References}{ref}
\LastPageEnding

\end{document}